\documentclass[final,5p,times,twocolumn,11pt,authoryear]{elsarticle}

\usepackage{graphicx}
\usepackage{amssymb}
\usepackage{amsmath}
\usepackage{empheq} 
\usepackage{comment}
\usepackage{enumerate}
\usepackage{setspace}
\usepackage[inline,shortlabels]{enumitem}

\usepackage{footmisc} 					
\usepackage{balance} 					
\usepackage[usenames,dvipsnames]{color} 	
\usepackage{floatpag} 					
\usepackage{bbm}						
\usepackage[normalem]{ulem} 				
\useunder{\uline}{\ul}{} 					

\numberwithin{equation}{section} 
\biboptions{sort&compress} 

\makeatletter
\def\@author#1{\g@addto@macro\elsauthors{\normalsize%
    \def\baselinestretch{1}%
    \upshape\authorsep#1\unskip\textsuperscript{%
      \ifx\@fnmark\@empty\else\unskip\sep\@fnmark\let\sep=,\fi
      \ifx\@corref\@empty\else\unskip\sep\@corref\let\sep=,\fi
      }%
    \def\authorsep{\unskip,\space}%
    \global\let\@fnmark\@empty
    \global\let\@corref\@empty  
    \global\let\sep\@empty}%
    \@eadauthor={#1}
}
\makeatother

\makeatletter
\def\ps@pprintTitle{%
 \let\@oddhead\@empty 
 \let\@evenhead\@empty
 \def\@oddfoot{}%
 \let\@evenfoot\@oddfoot}
 \makeatother

\usepackage[table,xcdraw,svgnames]{xcolor}
\usepackage[colorlinks]{hyperref}
\AtBeginDocument{%
  \hypersetup{
    citecolor=SteelBlue,
    linkcolor=SteelBlue,   
    urlcolor=SteelBlue}
   }
   
\journal{~}
\begin{document}
\begin{frontmatter}

\title{Ordinal Tax To Sustain a Digital Economy}

\author{Nate Dwyer}

\author{Sandro Claudio Lera\corref{cor1}}
\cortext[cor1]{corresponding author: Sandro Claudio Lera (slera@mit.edu)}

\author{Alex `Sandy' Pentland}

\address{Massachusetts Institute of Technology, 77 Massachusetts Avenue, 02139 Cambridge, Massachusetts, USA}

\begin{abstract}

Recently, the French Senate approved a law that imposes a $3\%$ tax on revenue generated from digital services by companies above a certain size. 
While there is a lot of political debate about economic consequences of this action, it is actually interesting to reverse the question: 
We consider the long-term implications of an economy with no such digital tax. 
More generally, we can think of digital services as a special case of products with low or zero cost of transportation. 
With basic economic models we show that a market with no transportation costs is prone to monopolization as minuscule, random differences in quality are rewarded disproportionally. 
We then propose a distance-based tax to counter-balance the tendencies of random centralisation.  
Unlike a tax that scales with physical (cardinal) distance, a ranked (ordinal) distance tax leverages the benefits of digitalization while maintaining a stable economy. 

\end{abstract} 

\end{frontmatter}

Until not too long ago, people used to live in relatively self-contained tribes or villages. 
Basic products such as bread and textiles were produced locally. 
Trade primarily existed for specialised products.  
Over time, from the industrial revolution through present technological advancements in AI, more efficient distribution channels have emerged.
Online orders are often delivered within a couple days, independent of a customer’s physical location. 
Extrapolated, this trend leads to an economy where, at least domestically, physical shipping distance becomes irrelevant. 
For digital products and services, this has largely become a reality and contradicts established economic theories, like \cite{Isard1954}'s gravity model of trade which states that trade activity is inversely proportional to physical distance. 

\begin{figure}[!htb]
	\centering
	\includegraphics[width=0.3\textwidth]{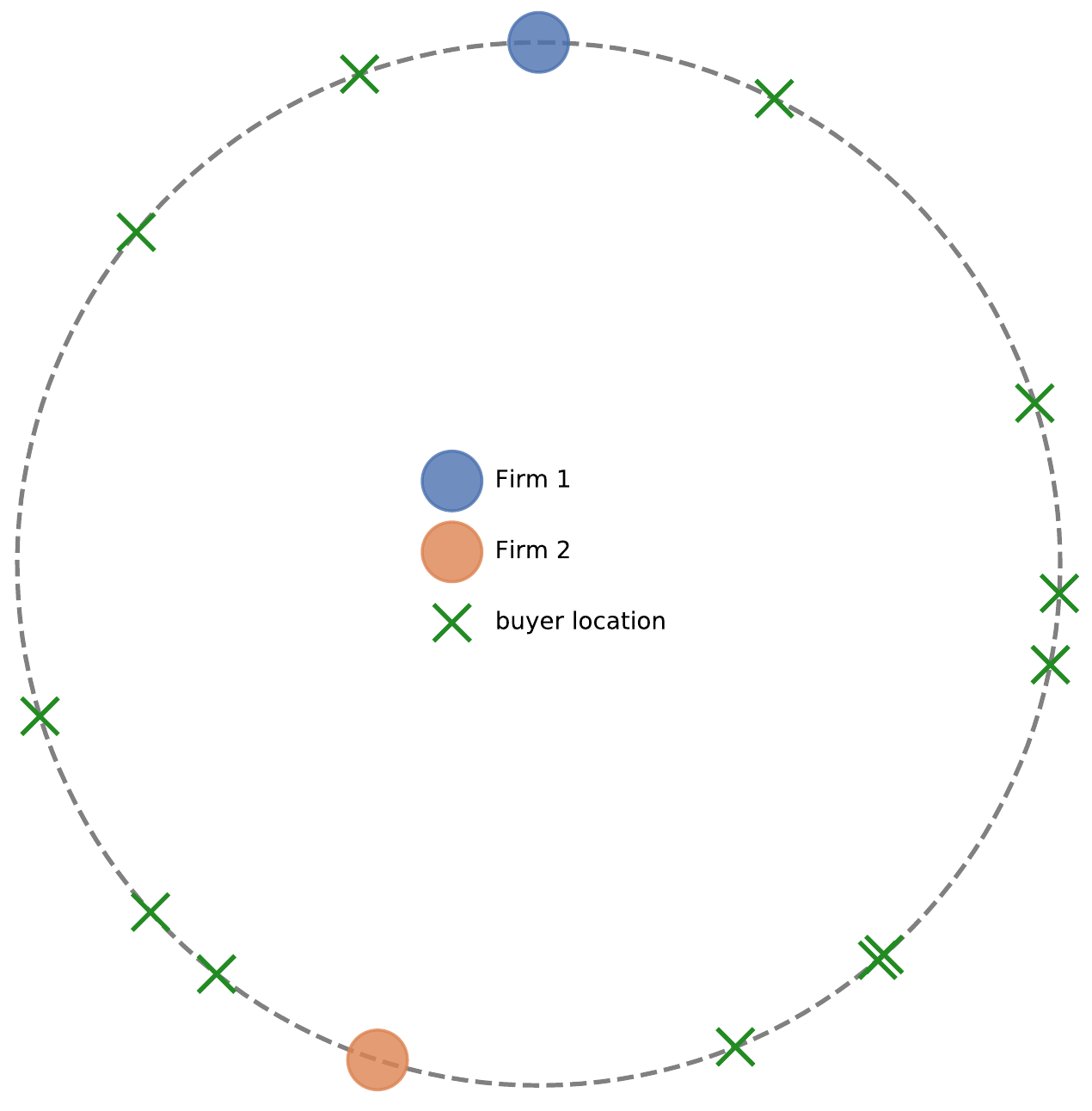}
	\caption{	Ring of length $\ell=1$ representing a `geography', on which we place at random two firms (circles) and $n=12$ buyers (crosses). 
			Here we have picked a balanced configuration where each firm has $n=6$ closest potential customers. 
			Throughout this article, we consider the buyer and firm arrangement fixed as shown here. 
			Any other generic arrangement or geography could be used and would not significantly alter our results.
			The two firms compete for profit.
			They set their production quantity and prices according to the Nash equilibrium, 
			taking into consideration the different levels of taxation. 
			}
	\label{fig:topology}
\end{figure}

To formally analyse the potential impact of vanishing distance-based transaction costs, we consider the following simple model: 
Two firms offering the same product are competing for a total of $n$ potential buyers. 
Firms and buyers are randomly placed on a one-dimensional ring of length $\ell$ with periodic boundary conditions (Figure \ref{fig:topology}). 
Hence, the maximum distance between any buyer and firm is $d=\ell/2$. 

Firm $i$ $(i=1,2)$ produces $q_i$ units of the good at unit cost $c_i$ and offers to sell at a price $p_i$. 
Their profit calculates to $\Pi_i = p_i \hat{q}_i - c_i q_i$ where $\hat{q}_i$ is the quantity that is actually sold ($\hat{q}_i \leqslant q_i$). 

\begin{figure}[!htb]
	\centering
	\includegraphics[width=0.5\textwidth]{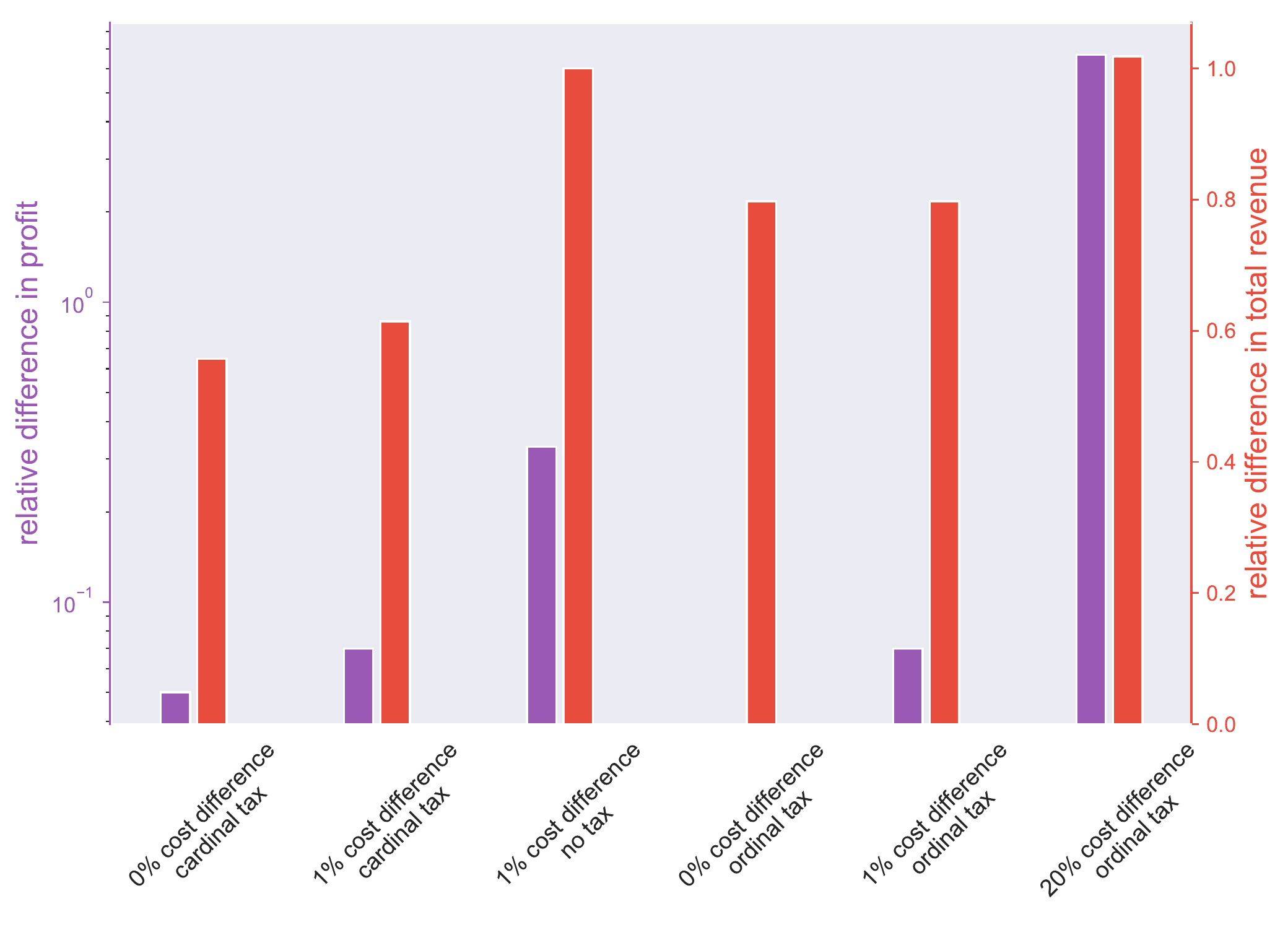}
	\caption{	On the left axis (blue bars), we show the relative difference in profit of the two firms.
			If both firms have equal production cost, their revenue is similar. 
			This remains true if there is a marginal $1\%$ difference in cost, as long as there is a distance dependent cardinal tax. 
			Imbalance in revenue is increased drastically once that tax is removed, but is decreased again in case it is replaced by an ordinal tax. 
			Major imbalances in production costs (i.e. competitive advantage) lead to significantly different revenue with and without tax. 
			On the right axis (orange bars), we show the relative difference in total market revenue, as generated by the two firms combined. 
			The revenue is rescaled to a value of $1$ for the case of no tax. 
			All other revenue values are in proportion thereof. 
			While the cardinal tax significantly decreases revenue, the ordinal tax brings it back to a similar level than without any tax. 
			}
	\label{fig:results}
\end{figure}

The buyers cannot buy at price $p_i$, but additionally have to pay a transaction cost $\lambda \cdot d^\gamma \cdot p_i$ where $d$ is the distance between a given buyer and firm $i$. 
From hereon, we shall call this the \textit{cardinal tax}. 
In its most basic form, the gravity model predicts a value of $\gamma=2$, whereas here we have set $\gamma=1$.  
\footnote{
\label{ftn:wlog}
Changing these numbers does not alter the qualitative statements of our results, 
while some of the numbers quoted in Figure \ref{fig:results} may change.
} 
The case of zero cost of transportation is recovered in the limit where the scaling factor $\lambda$ approaches zero. 
The buyers are all characterised by the same negatively sloped demand curve, $p = u - q$ with $u$ the upper price limit at which no good is bought.
In the remainder of this article, we set $\lambda = 0.1$ and $u = 120$. \footref{ftn:wlog}

The two firms choose capacity and price to maximize their respective profit in equilibrium. 
Here, we assume they first compete on capacity, by simultaneously declaring how many units they produce;
next, they compete on price, taking into consideration the tax and assuming the location of buyers is known. 
This process, consisting of these two dependent sub-games, is similar to the game theoretical problems studied by \cite{Kreps1983}, the major difference being the buyer-seller specific tax.
Quantity and price offered by each firm are determined as the Nash equilibrium of this two-stage game. 
Because the tax depends on the exact location of buyers and firms, 
it is not possible to determine the Nash equilibrium analytically, for general cases. 
Instead, we use the Gambit library developed by \cite{McKelvey2016} to find the Nash equilibrium numerically. 

As both firms are offering the same good, they differ only in their marginal cost of production $c_i$ (and their physical location, as shown in Figure \ref{fig:topology}). 
Let us consider the case where both firms are equally competitive ($c_1 = c_2 = 100$), solve for Nash equilibrium prices and consider their profits. 
We find that their relative difference in profit is $5\%$ (see left axis in Figure \ref{fig:results} for a visualisation of this and subsequent results).
This difference is relatively small as we have chosen a rather homogeneous distribution of buyers and firms (Figure \ref{fig:topology}).
If the distribution of buyers were more irregular, so would be the profits, and the firms would ultimately move to similar locations, as was already predicted by \cite{Hotelling1929}.

If there is no cost of transportation, the difference in profit is zero as predicted by symmetry (both firms are exactly equal).
However, this equality is highly unstable. 
To see this, let us now consider the case where firm $1$ is marginally more competitive ($c_1 = 99, c_2 =100$). 
As long as there is a cost of transportation $(\lambda=0.1)$, the difference in profit merely increases from $5\%$ to $7\%$.
But in absence of such a cardinal tax ($\lambda=0$), the difference in profit amounts to a whopping $33\%$!
This result is particularly bothersome as this excess profit can be used to further decrease costs of production.
Either through genuine innovation, or temporary undercutting of prices, the difference between $c_1$ and $c_2$ is further enlarged and the second firm is eventually run out of business. 
\footnote{
This claim can be formalised by considering multi-stage games.
} 
In general, the outpace of competition is natural and in many cases even desired, as efficiency is to be rewarded. 
However, we stress that a minor, essentially random difference of $1\%$ production cost disproportionally benefited one firm. 
We may rationalise this by going back to the example of self-contained villages from the introduction: 
Once there is no more cost of transportation, the economic agent has little incentive to favour local bread, if bread from another village is just marginally cheaper. 
Monopolisation initiated by minor, random initial differences in quality is thus to be expected. 

These ever decreasing (or non-existing) cost of transportation can be counter-balanced by introducing a tax that scales with distance (e.g. a carbon tax). 
An immediate side-effect of such market interference is less consumption and hence decreased revenue for both firms. (second vs. third bar on right axis in Figure \ref{fig:results}). 
While decentralised, locally self-contained economies may benefit consumers on a grand scheme, it certainly punishes remote buyers and is susceptible to potentially small changes in location. 
Therefore, we propose to introduce an \textit{ordinal tax} instead. 
For each buyer, the distance to the different firms is enumerated in increasing order. 
A buyer has a distance of $D=0$ from the physically (cardinally) closest firm, a distance $D=1$ from the second closest, and so forth. 
The advantages of this tax are manifold. 
First, it resolves the issue of remote buyers being worse off. 
Second, it is not susceptible to (small) changes in buyer or firm locations, and hence does not lead to physical centralisation (profits are exactly equal if $c_1=c_2$). 
Third, the overall level of revenue is elevated compared to a cardinal tax, as now only remote purchases $(D > 0)$ are taxed (right axis in Figure \ref{fig:results}). 
Fourth, marginal differences in quality remain unnoticed, as the difference in production cost does not outweigh taxation costs. 
Fifth, major difference in quality (true competitive advantage), overcomes the tax and rewards the more competitive firm 
(cf. in Figure \ref{fig:results} the $50$ times higher revenue in case of a $20\%$ difference in production cost). 

The ordinal tax is similar in spirit to, but more dynamic than a country border tax. 
Most importantly, specialised products are traded without tax ($D=0$ everywhere) such that innovation spreads globally. 
Products that are easy to produce remain local. 
Competition is encouraged and independent of a firms location, so there is no a priori protectionism. 
However, to actually outperform, significant competitive advantages must be present. 
As such, the ordinal tax acts as a type of centralisation threshold that must be overcome. 
The value of that threshold is tuned by the scaling factor $\lambda$. 
One can interpolate between the two extremes of no transportation tax $(\lambda=0)$ and forced local monopolisation $(\lambda=\infty)$. 
A self-regulating adjustment of $\lambda$ based on different market conditions, targeted redistribution of tax money and a practical implementation of this tax on digital markets are part of ongoing research.

\section*{Acknowledgements} 

S. Lera acknowledges stimulating discussions with S. Donick.

\bibliographystyle{apa}
\bibliography{bibliography.bbl}

\begin{thebibliography}{}

\bibitem[\protect\astroncite{Hotelling}{1929}]{Hotelling1929}
Hotelling, H. (1929).
\newblock {Stability in Competition}.
\newblock {\em The Economic Journal}, 39(153):41--57.

\bibitem[\protect\astroncite{Isard}{1954}]{Isard1954}
Isard, W. (1954).
\newblock {Location theory and trade theory: short-run analysis}.
\newblock {\em The Quarterly Journal of Economics}, pages 305--320.

\bibitem[\protect\astroncite{Kreps and Scheinkman}{1983}]{Kreps1983}
Kreps, D.~M. and Scheinkman, J.~A. (1983).
\newblock {Quantity precommitment and Bertrand competition yield Cournot
  outcomes}.
\newblock {\em The Bell Journal of Economics}, pages 326--337.

\bibitem[\protect\astroncite{McKelvey and Turocy}{2016}]{McKelvey2016}
McKelvey, R.~D., M. A.~M. and Turocy, T.~L. (2016).
\newblock {Gambit: Software Tools for Game Theory, Version 16.0.1}.

\end{thebibliography}

\end{document}